\documentclass[a4paper,11pt]{article}
\pdfoutput=1 % if your are submitting a pdflatex (i.e. if you have
\usepackage{jcappub} % for details on the use of the package, please
\usepackage[T1]{fontenc} % if needed
\usepackage{hyperref}

\title{\boldmath Charged gravastar model in Rastall theory of gravity}

\author[1]{Debadri Bhattacharjee}
\author[2]{and Pradip Kumar Chattopadhyay}
\affiliation[1,2]{IUCAA Centre for Astronomy Research and Development (ICARD), Department of Physics, Cooch Behar Panchanan Barma University, Vivekananda Street, District: Cooch Behar, \\ Pin: 736101, West Bengal, India}
\emailAdd{debadriwork@gmail.com}
\emailAdd{ pkc$_{-}$76@rediffmail.com}
\abstract{Gravastars are considered prime exotic compact objects that may be found at the end of gravitational collapse of massive stars to resolve the complexities that are common in the case of black holes \cite{Mazur}-\cite{Mazur2}. In this paper, we analyse the role of charge on the possible formation of an isotropic spherically symmetric gravastar configuration in the framework of Rastall gravity. Gravastar contains three distinct layers {\it viz.} i) Interior region, ii) Thin shell and iii) Exterior region. The interior region is characterised by the equation of state, $p=-\rho$, which defines the repulsive outward pressure in the radial direction at all points on the thin shell. The thin shell, contains an ultra-relativistic stiff fluid, which is denoted by the equation of state $p=\rho$ following Zel'dovich's criteria \cite{Zeldovich,Zeldovich1} for a cold baryonic universe, and can withstand the repulsive pressure exerted by the interior region. The exterior region is the vacuum space-time represented by the Reissner-Nordstr$\ddot{o}$m solution. In view of the above specifications, we construct and analyse a charged gravastar model in the framework of Rastall theory of gravity, which represents several salient features. The basic physical attributes, {\it viz.} proper length, energy, entropy and equation of state parameter of the shell are investigated. In this model, it is interesting to note that for a large radius of the hypersurface $(R)$, the equation of state parameter of the thin shell corresponds to a dark energy equation of state with $\mathcal{W}(R)\rightarrow-1$. However, for a small value of $R$, $\mathcal{W}(R)\rightarrow0$, defines a dust shell. The stability of the model is ensured through the study of gravitational surface redshift and maximisation of shell entropy within the framework of Rastall theory of gravity.}

\begin{document}
\maketitle
\flushbottom

\section{Introduction}\label{sec1} Compact astrophysical objects mark the end state of stellar evolution. When a stellar body is unable to counterbalance its own gravitational collapse though the generation of thermal pressure, a compact object is created. The theory and evolution of compact objects have attracted the attention of astrophysicists, as this terminal point of stellar collapse creates extreme pressure and density regimes to test different relevant theories within the context of the general theory of relativity. It is conjectured that if a massive stellar body goes through a supernova explosion, then, depending on the mass of the remnant, it ends either as a neutron star or a black hole. In 1916, Schwarzschild provided the first black hole solution from the Einstein Field Equation (EFE) in vacuum, and this solution is termed the most important solution for a classical charged as well as uncharged black hole. However, the Schwarzschild solution has two fundamental drawbacks, {\it viz.}, i) the existence of central singularity and ii) the presence of an event horizon. The central singularity $(r\rightarrow0)$ is an irremovable singularity that cannot be removed by a certain coordinate transformation; this singularity is called a dynamical singularity. However, the singularity at the Schwarzschild radius $r=r_{s}=\frac{2GM}{c^{2}r}$, which in the case of a black hole constitutes the event horizon at $r=2M$ (in a system of units $G=c=1$), is termed the kinematic or coordinate singularity and poses fundamental problematic situations. In terms of quantum mechanics, the presence of an event horizon shows that near this region, the energy of a photon diverges, and there are no a-priori parameters available to control this deformity in the region. We know that, the Hawking temperature is inversely proportional to the mass of the black hole $(T=\frac{\hbar}{8\pi k_{B}GM})$ \cite{Hawking}, which in turn shows that black holes in thermal equilibrium with their own Hawking radiation have negative specific heat, which makes them unstable under thermodynamic fluctuations. Again, black holes have a gigantic amount of entropy and the presence of information paradox, which is another viable condition for seeking an alternative approach. Mazur and Mottola \cite{Mazur}-\cite{Mazur2} extended the idea of the gravitational Bose-Einstein condensate (GBEC) to hypothesise a new model of cold, dark compact objects named Gravitational Vacuum Stars or abbreviated as Gravastars. The basic model of gravastars contained five layers, which were subsequently reduced to a three-layer model by Visser and Wiltshire \cite{Visser}. It is now widely accepted that gravastars have a three-layered structure, namely, an interior region, an exterior region and an intermediate thin shell separating the interior and exterior spaces. Three regions are specified by three distinct equation of states (EoS). The interior region being in de-Sitter condensate phase is characterised by positive pressure and negative energy density $(p=-\rho)$, which is equivalent to the positive cosmological constant that Einstein proposed to modify the static universe model based on Mach's principle. The presence of negative energy density shows that the interior region exerts repulsive outward pressure at all points on the thin shell. The thin shell is hypothesised to contain an ultra-relativistic stiff fluid with an EoS $p=\rho$, which is consistent with Zel'dovich criterion of the EoS in the context of a cold baryonic universe. Within the context of causality, the velocity of sound and the velocity of light are equal in this region. The shell is thin finite in thickness. The exterior region is a vacuum flat space-time region, as indicated by the EoS $p=\rho=0$. 

In their work, Mazur and Mottola \cite{Mazur}-\cite{Mazur2} hypothesised that gravastars are an alternative manifestation of black holes, and they studied the stability of the model through the concept of entropy maximisation. Visser and Wiltshire \cite{Visser} investigated the dynamical stability of the gravastar model against spherically symmetric perturbations of matter or the gravity field. In pursuing stability, they developed a mathematical structure, compatible with the Mazur-Mottola hypothesis, that describes the stability of the model through realistic values of the EoS parameter. Cattoen et al. \cite{Cattoen} imposed pressure anisotropy on the formation of gravastars. They claimed that gravastars cannot be perfect fluids and that the inclusion of pressure anisotropy could be useful in supporting gravastars with high compactness. Considering the range of associated parameters, Carter \cite{Carter} investigated the existence of a thin shell and the stability of gravastars with Reissner-Nordstr\"om \cite{Reissner,Nordstrom} exterior. He also explored the dependence of the EoS parameter on possible gravastar formation. Regarding the theoretical modeling of gravastars in the presence of an electromagnetic field, Horvat et al. \cite{Horvat} introduced two models. Using suitable matching conditions, they obtained stable gravastars through dominant energy conditions. In their approach, they studied the effects of electromagnetic field on the EoS parameter, the velocity of sound and surface redshift. In the framework of conformal motion, Usmani et al. \cite{Usmani} proposed a charged gravastar model considering the Reissner-Nordstr\"om exterior, and the model was further generalised to a higher dimensional approach by Bhar \cite{Bhar}. Rahaman et al. \cite{Rahaman} studied the viable configuration of charged gravastars in (2+1)-dimensional space-time. In a new approach, Bhattacharjee and Chattopadhyay \cite{Bhattacharjee} studied the formation of charged gravastars in a generalised cylindrically symmetric space-time. Considering the charged shell configuration, Chan and Silva \cite{Chan} studied the effect of charge on the stability of charged gravastars. In the framework of of non-commutative geometry, \"Ovg\"un et al. \cite{Ovgun} constructed charged thin shell gravastars.
 
An interesting approach to understand the structure formation and evolution of the universe is through the light of modified gravity theories. Although Einstein's general theory of relativity is the most promising theory for revealing the mysteries of the universe, observational evidence in favour of the notion of an accelerating universe, dark energy and dark matter poses serious challenges to the theoretical predictions of general relativity \cite{Riess}-\cite{Tegmark}. The importance and need for incorporating modified gravity have been elaborately discussed in \cite{Nojiri}. The generalisation of the Einstein-Hilbert action leads to the formalism of various alternative theories of gravity, {\it viz.} $f(R)$ ($R$ is the Ricci scalar) \cite{Capozziello,Elizalde}, $f(\mathcal{T})$ ($\mathcal{T}$ is the torsion scalar) \cite{Bamba}, $f(R,\Box R,T)$ ($\Box$ is the de Alembert's operator, and $T$ is the trace of the energy-momentum tensor) \cite{Houndjo}-\cite{Ilyas},$f(R,T)$ \cite{Harko}, etc. Many studies have been performed on gravastars in the skeleton of modified gravity theories \cite{Das}-\cite{Banerjee}. In comparison to the general theory of relativity, the interesting feature of these theories is the non-conservation of the energy-momentum tensor which leads to the non-minimal coupling of geometry and matter fields. This intriguing aspect of null covariant divergence is the main motivation for us to model a charged gravastar system in Rastall theory of gravity, as introduced by P. Rastall \cite{Rastall,Rastall1}. In this approach, the covariant derivative of the energy-momentum tensor is proportional to the divergence of the Ricci scalar curvature, which is expressed as $\nabla_{\nu}T^{\nu}_{\mu}=\eta \mathfrak{R}_{,~\mu}$. 

The Rastall theory provides a straightforward and tractable form for the EFE that yields interesting features from astrophysical and cosmological perspectives. In the recent past, Rastall theory has been found to be successful in the formalism of black hole solutions \cite{Heydarzade}, charged and rotating NUT black holes \cite{Parihadi}, rotating blak holes \cite{Kumar}, Kerr-Newman-AdS black holes \cite{Xu}, black holes inspired by non-commutative geometry \cite{Ma}, nonlinear charged black holes \cite{Gergess} etc. The Rastall theory has also been important in compact star modeling. In a general approach, Mota et al. \cite{Mota} described the effects of Rastall gravity on compact astrophysical objects. Using the Krori and Barua metric ansatz \cite{KB}, Abbas and Shahzad \cite{Abbas} studied the anisotropic solutions of compact stars in Rastall theory. In another work, Abbas and Shahzad \cite{Abbas1} explored the same formalism by varying the cosmological constant. Neutron stars in the framework of Rastall gravity were studied by Oliveira et al. \cite{Oliveira}. The effect of Rastall gravity on the basic characteristic features, i.e., mass, radius, sound velocity etc. of PSR J0740+6620 was studied in Ref. \cite{Hanafy}. Considering perfect fluid spheres, the impact of Rastall parameter was investigated in Ref. \cite{Hansraj}. In the literature, there are numerous studies that depict the impact of Rastall theory on the astrophysical and cosmological aspects \cite{Capone}-\cite{Mustafa}.  

The remainder of this paper is organised as follows. In section \ref{sec2}, we set up the Einstein-Maxwell field equations along with the notion of non-conservation of the energy-momentum tensor in the framework of Rastall gravity. In section \ref{sec3}, we derive the solutions of the field equations in the three distinct layers of gravastar, {\it viz.}, interior region, thin shell and exterior vacuum space-time for the characteristic EoS of the three regions. In this section, we have also obtained the expression for the pressure and matter density of the thin shell. The modification of junction condition in Rastall theory and smooth matching of the interior and exterior solutions at the hyper-surface $(r=R)$ is ensured in section \ref{sec4}. Section \ref{sec5} addresses with the boundary conditions, i.e., the matching of the interior and thin shell solutions at the interior boundary $(r=r_{1})$ and thin shell and exterior solutions at the exterior boundary $(r=r_{2})$, to determine the numerical values of the necessary constants. Several fundamental characteristics of a gravastar, namely, proper length, energy and entropy of the thin shell, the EoS parameter of the hyper-surface and the impact of charge on these attributes are studied in the section \ref{sec6}. In section \ref{sec7}, we have analysed the stability of the model on the basis of the surface redshift parameter $(Z_{s})$ and the consideration of entropy maximisation. Finally in section \ref{sec8} we have discussed the major findings of the paper.       
\section{The Field Equations in Rastall Gravity}\label{sec2}
We begin with a short introduction to the Rastall theory as introduced by P. Rastall \cite{Rastall,Rastall1}. In the original idea \cite{Rastall1}, Rastall modified the Einstein theory of gravity by considering the non-vanishing divergence of the energy-momentum tensor, which is proportional to the covariant derivative of the Ricci scalar expressed as:  
\begin{equation}
	\nabla_{\nu}T^{\nu}_{\mu}=\beta_{\mu}, \label{eq1}
\end{equation}
where $\beta$ sustains its form in curved space-time but vanishes in a flat geometry. We note that $\beta_{\mu}=\eta \mathfrak{R}_{,~\mu}$, where $\eta$ is a constant, $\mathfrak{R}=g^{\mu\nu}\mathfrak{R_{\mu\nu}}$ is the Ricci Scalar and $T^{\nu}_{\mu}$ is the energy-momentum tensor expressed as $T^{\nu}_{\mu}=(T^{\nu}_{\mu})_{m}+(T^{\nu}_{\mu})_{q}$, where the subscripts define the matter and electromagnetic (charge) sectors respectively. In the context of Rastall theory, the consideration of a non-minimal coupling between the matter field and geometry leads to the modification of Einstein's tensor written as:
\begin{equation}
	G_{\mu\nu}+k\eta g_{\mu\nu} \mathfrak{R}=kT_{\mu\nu}. \label{eq2}
\end{equation}  
The relation connecting the Ricci scalar and energy-momentum tensor is modified as:
\begin{equation}
	\mathfrak{R_{\mu\nu}}+(k\eta-\frac{1}{2})g_{\mu\nu}\mathfrak{R}=kT_{\mu\nu}. \label{eq3}
\end{equation}
In Rastall gravity, $k$ is the gravitational coupling parameter. By contracting \eqref{eq3}, one obtains, $\mathfrak{R}=\frac{kT_{\mu}^{\mu}}{4k\eta-1}$. In the present case, we exclude $k\eta=\frac{1}{4}$ as this leads to $T_{\mu}^{\mu}=0$ and $\mathfrak{R}=0$, which in turn reduces to Einstein theory with $\mathfrak{R}_{\mu\nu}=0$. We consider $\zeta=k\eta$ as the dimensionless Rastall parameter. Therefore, in the units of $G=c=1$, the gravitational coupling parameter $k$ and constant $\eta$ are expressed as:
\begin{equation}
	k=8\pi\frac{4\zeta-1}{6\zeta-1}, \label{eq4}
\end{equation}  
and 
\begin{equation}
	\eta=\frac{\zeta(6\zeta-1)}{8\pi(4\zeta-1)}. \label{eq5}
\end{equation}
It is evident from Eq.~\eqref{eq4} that for $\zeta=0$ $k$ reduces to the Einstein gravitational constant. From Eqs.~\eqref{eq4} and \eqref{eq5}, we can see that for $\zeta=\frac{1}{6}$, $k$ diverges, and again, we exclude $\zeta=\frac{1}{4}$ for Rastall gravity properties. In relativistic units, $G=c=1$, the field equations in Rastall theory of gravity take the form:
\begin{equation}
		G_{\mu\nu}+\zeta g_{\mu\nu} \mathfrak{R}=8\pi T_{\mu\nu}\Big(\frac{4\zeta-1}{6\zeta-1}\Big). \label{eq6}
\end{equation}    
The static spherically symmetric space-time is expressed as: 
\begin{equation}
	ds^2=-e^{2\nu(r)}dt^2+e^{2\lambda(r)}dr^2+r^2(d\theta^2+sin^2\theta d\phi^2). \label{eq7}
\end{equation}
In eq.~\eqref{eq6}, $T_{\mu\nu}=(T_{\mu\nu})_{m}+(T_{\mu\nu})_{q}$, is the composite energy-momentum tensor, where the subscripts depict the matter sector and electromagnetic(charge) distribution, respectively. The perfect fluid consideration of the interior matter sector is expressed in the following form:
\begin{equation}
	(T_{\mu\nu})_{m}=(p+\rho)u_{\mu}u_{\nu}+pg_{\mu\nu}, \label{eq9}
\end{equation}
where  $\rho$ is the energy density, $p$ is the isotropic pressure and $u^{\mu}~(u^{\mu}u_{\mu}=-1)$ is the four-velocity of the perfect fluid in consideration. On the other hand, the electromagnetic(charge) distribution takes the form:
\begin{equation}
	(T_{\mu\nu})_{q}=\frac{1}{4\pi}(F^{\sigma}_{\mu}F_{\nu\sigma}-\frac{1}{4}g_{\mu\nu}F_{\sigma\beta}F^{\sigma\beta}). \label{eq10}
\end{equation}
Here, $F_{\mu\nu}$ is the Faraday-Maxwell tensor defined by the form, $F_{\mu\nu}=\partial_{\mu}{A_{\nu}}-\partial_{\nu}{A_{\mu}}$, where $A_{\mu\nu}$ is the electromagnetic four potential. The corresponding Maxwell electromagnetic field equations are
\begin{equation}
	(\sqrt{-g}F^{\mu\nu})_{,\nu}=4\pi J^{\mu}\sqrt{-g}~~, ~~~~F_{[\mu\nu,\delta]}=0. \label{eq11}
\end{equation}
Here, $J^{i}$ is the electric four current obeying $J^{i}=\sigma u^{i}$ where $\sigma$ is the charge density and the corresponding electric field intensity $E$ is defined through the charge parameter $q$ as, 
\begin{equation}
	E=\frac{q}{4\pi r^2}. \label{eq12}
\end{equation}
Within the framework of Rastall gravity, eqs.~\eqref{eq6}-\eqref{eq10} fabricate the following set of tractable field equations: 
\begin{equation}
	\frac{2\lambda'e^{-2\lambda}}{r}+\frac{(1-e^{-2\lambda})}{r^2}+\zeta e^{-2\lambda}\Big[2\nu''+2(\nu')^2-2\lambda'\nu'+\frac{4}{r}(\nu'-\lambda')+\frac{2}{r^2}(1-e^{2\lambda})\Big]=\frac{(4\zeta-1)}{6\zeta-1}(8\pi\rho+\frac{q^2}{r^4}),\label{eq13}
\end{equation}\\
\begin{equation}
	\frac{2\nu'e^{-2\lambda}}{r}-\frac{(1-e^{-2\lambda})}{r^2}-\zeta e^{-2\lambda}\Big[2\nu''+2(\nu')^2-2\lambda'\nu'+\frac{4}{r}(\nu'-\lambda')+\frac{2}{r^2}(1-e^{2\lambda})\Big]=\frac{(4\zeta-1)}{6\zeta-1}(8\pi p-\frac{q^2}{r^4}),\label{eq14}
\end{equation}\\
\begin{eqnarray}
e^{-2\lambda}\Big[\nu''+(\nu'^2)-\lambda'\nu'+\frac{\nu'-\lambda'}{r}\Big]-\zeta e^{-2\lambda}\Big[2\nu''+2(\nu')^2-2\lambda'\nu'+\frac{4}{r}(\nu'-\lambda')+\frac{2}{r^2}(1-e^{2\lambda})\Big]\nonumber\\=\frac{(4\zeta-1)}{6\zeta-1}(8\pi p+\frac{q^2}{r^4}).\nonumber\\\label{eq15}
\end{eqnarray}
Here, $\lq\prime$' denotes the differentiation with respect to the radial parameter $r$, and we have considered the formalism in the geometrised unit of $G=c=1$. The conservation of the energy-momentum tensor in Rastall theory leads to
\begin{equation}
	p'+(p+\rho)\nu'-\frac{\zeta}{4\zeta-1}(3p'-\rho')=0. \label{eq16} 
\end{equation}
It is interesting to note that, in case of $\zeta\rightarrow0$ one retains the Einstein theory of gravity. 
\section{Gravastar Model}\label{sec3}
In the preceding section, we study the structure of gravastar in Rastall gravity regime. 
\subsection{Interior Region} The interior region of a gravastar has been hypothesised to be a vacuum de-Siiter space characterised by the EoS, $p=-\rho$ \cite{Mazur}-\cite{Mazur2}. The negative or repulsive pressure in this region is responsible for the radial outward force on all points on the shell, which counterbalances the inward gravitational collapse. On the other hand, the negative energy density in this region is equivalent to the positive cosmological constant. This particular choice of EoS is also termed the \lq degenerate vacuum' or \lq$\rho$-vacuum' and it is a prime example of a dark energy EoS. Using the EoS in eq.~\eqref{eq16}, we obtain, $p=-\rho=-\rho_{c}$, where $\rho_{c}$ is a constant implying a constant interior density. Applying the EoS in eqs.~\eqref{eq13} to \eqref{eq15}, we obtain the form of metric potential $\lambda(r)$ as:
\begin{equation}
	e^{-2\lambda(r)}=1+\frac{1}{3}c_{1}r^2-\frac{k}{8\pi}\frac{q^2}{r^2}+\frac{c_{2}}{r}, \label{eq17}
\end{equation}  
where $c_{1},~ c_{2}$ are the integration constants. To ensure the regularity of the solution at the center, one can demand $c_2=0$. Therefore, eq.~\eqref{eq17} reduces to 
\begin{equation}
		e^{-2\lambda(r)}=1+\frac{1}{3}c_{1}r^2-\frac{k}{8\pi}\frac{q^2}{r^2}. \label{eq18}
\end{equation}
From eqs.~\eqref{eq14} and \eqref{eq15} we obtain:
\begin{equation}
	e^{2\nu(r)}=c_{3}\Big(1+\frac{1}{3}c_{1}r^2-\frac{k}{8\pi}\frac{q^2}{r^2}\Big), \label{eq19}
\end{equation}
where $c_{3}$ is an integration constant. It should be noted that eqs.~\eqref{eq18} and \eqref{eq19} are non-singular solutions. Therefore, the problem of central singularity in the interior region of the gravastar is removed. \\
The active gravitational mass of the interior regions is obtained as:
\begin{equation}
	M_{active}=\int_{0}^{r_{1}}(\rho+\frac{q^{2}}{8\pi r^4})dr=\frac{4\pi\rho_{c}r_{1}^{3}}{3}-\frac{q^2}{2r_{1}}. \label{eq20}
\end{equation}
\subsection{Thin Shell} 
\begin{figure}[htbp]
	\centering
	\includegraphics[width=.6\textwidth]{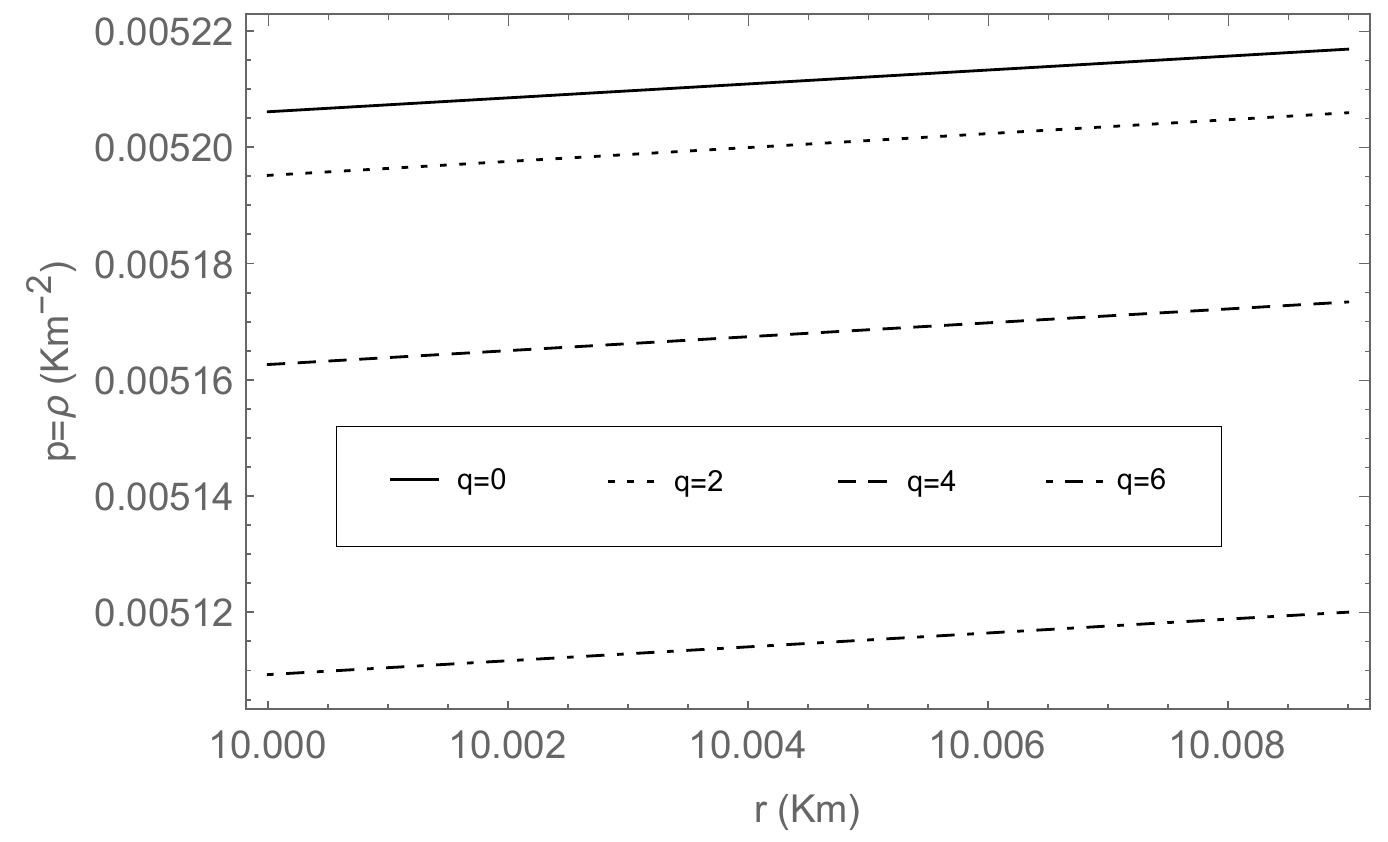}
	\caption{Variation of matter density or pressure with radius r(km).\label{fig1}}
\end{figure}

The intermediate thin shell contains ultra-relativistic fluid obeying the EoS $p=\rho$. Against the backdrop of the cold baryonic universe, Zel'dovich \cite{Zeldovich,Zeldovich1} proposed the idea of this type of fluid, which is also termed as \lq stiff fluid'. In the present work, we may argue that this kind of approximation may be taken into consideration due to the thermal fluctuations with zero chemical potential or due to the conservation of the number density of gravitational quanta at zero temperature. Several researchers \cite{Madsen}-\cite{Braje} have used this stiff fluid to investigate different cosmological and astrophysical scenarios. The non-vacuum thin shell approximation along with the stiff fluid EoS make it a difficult to obtain the exact solutions for this region. However, analytical solutions can be obtained using the thin shell approximation, i.e., $0<e^{-2\lambda}\ll1$. Following the thin shell approximation along with the EoS and eqs.~\eqref{eq13}-\eqref{eq15}, we can obtain the following set of equations:
\begin{equation}
	\frac{2\lambda'e^{-2\lambda}}{r}+\frac{1}{r^2}+\zeta e^{-2\lambda}\Big[-2\lambda'\nu'-\frac{4\lambda'}{r}\Big]-\frac{2\zeta}{r^2}=\frac{(4\zeta-1)}{6\zeta-1}(8\pi\rho+\frac{q^2}{r^4}),\label{eq21}
\end{equation}
\begin{equation}
	-\frac{1}{r^2}-\zeta e^{-2\lambda}\Big[-2\lambda'\nu'-\frac{4\lambda'}{r}\Big]+\frac{2\zeta}{r^2}=\frac{(4\zeta-1)}{6\zeta-1}(8\pi p-\frac{q^2}{r^4}),\label{eq22}
\end{equation} 
\begin{equation}
	e^{-2\lambda}\Big[-\lambda'\nu'-\frac{\lambda'}{r}\Big]-\zeta e^{-2\lambda}\Big[-2\lambda'\nu'-\frac{4\lambda'}{r}\Big]+\frac{2\zeta}{r^2}=\frac{(4\zeta-1)}{6\zeta-1}(8\pi p+\frac{q^2}{r^4}).\label{eq23}
\end{equation}
From eqs.~\eqref{eq22} and \eqref{eq23} we obtain:
\begin{equation}
	\nu'=-\frac{1}{r}-\frac{e^{2\lambda}(r^2(1-6\zeta)+q^2(8\zeta-2))}{\lambda'r^4(6\zeta-1)}. \label{eq24}
\end{equation}
Using EoS $p=\rho$ in eqs.~\eqref{eq21},\eqref{eq22} and \eqref{eq24} we obtain:
\begin{equation}
	e^{-2\lambda}=\frac{(4\zeta-1)^2}{(6\zeta-1)(2\zeta-1)}\frac{q^2}{r^2}+2\frac{(4\zeta-1)}{2\zeta-1}\ln(r)+c_{4}. \label{eq25}
\end{equation}
Using eq.~\eqref{eq25} in eq.~\eqref{eq24} we obtain: 
\begin{equation}
	\nu=\frac{(6-16\zeta)\ln(r)+(2\zeta-1)\ln[q^2(1-4\zeta)+r^2(6\zeta-1)]}{(8\zeta-2)}+c_{5}, \label{eq26}
\end{equation}
where $c_{4}$ and $c_{5}$ are integration constants. Using eq.~\eqref{eq26} in eq.~\eqref{eq16} we obtain the matter density and pressure of the thin shell as:
\begin{equation}
	p=\rho=\rho_{0}e^{\frac{2(8\zeta-3)\ln(r)+(1-2\zeta)\ln{[q^2(1-4\zeta)+r^2(6\zeta-1)]}}{(2\zeta-1)}}. \label{eq26a}
\end{equation} 
The variation of the matter density as well as pressure with the charge and shell thickness are shown in figure~\ref{fig1}. It is to be noted here that the matter density or pressure increases toward the outer edge of the shell and decreases with increasing charge. 
\subsection{Exterior Region} The exterior region of the gravastar obeys the EoS $p=\rho=0$, which describes a vacuum space-time. Coupling the EoS with eqs.~\eqref{eq13} and \eqref{eq14}, we obtain:
\begin{equation}
	\lambda'+\nu'=0. \label{eq27}
\end{equation}
In the above consideration, we use the line element for the exterior region to be designated by the well known Reissner-Nordstr$\ddot{o}$m \cite{Reissner,Nordstrom} metric expressed as:
\begin{equation}
	ds^2=-\Big(1-\frac{2M}{r}+\frac{q^2}{r^2}\Big)dt^2+\Big(1-\frac{2M}{r}+\frac{q^2}{r^2}\Big)^{-1}dr^2+r^2(d\theta^2+sin^2\theta d\phi^2), \label{eq28}
\end{equation} 
where $M$ is the total mass and $q$ is the charge of the gravitating system. 
\section{Junction Condition}\label{sec4} The structure of a charged gravastar contains three layers. The interior, intermediate thin shell and the exterior region where the thin shell separates the the interior and exterior geometry. According to the fundamental junction condition, the smooth matching of the interior and exterior metrics at the hypersurface $r=R$ is ensured by the Israel criterion \cite{Israel,Israel1}. It should be noted here that even though the metric coefficients are continuous at the interface, their derivatives are not necessarily continuous. We consider the Darmois-Israel junction condition \cite{Israel}-\cite{Darmois} to compute the surface stresses at the junction interface. In Einstein gravity, Lanczos equation \cite{Lanczos}-\cite{Musgrave} in coordinates $X^{\alpha}=(t,r,\theta,\phi)$ expresses the intrinsic surface energy tensor $S_{\gamma\beta}$ in the following form:  
\begin{equation}
	{S}^{\gamma}_{\beta}=-\frac{1}{8\pi} \Big[{K}^{\gamma}_{\beta}-{\delta}^{\gamma}_{\beta}{K}^{\kappa}_{\kappa}\Big], \label{eq29}
\end{equation}
Following Ref. \cite{Rosa}, to describe the junction condition within the framework of Rastall theory of gravity, the distribution formalism for the Ricci tensor, Ricci scalar and energy-momentum tensor is expressed as: 
\begin{eqnarray}
	\mathfrak{R}_{\mu\nu}=\mathfrak{R}_{\mu\nu}^{\pm}-\delta(\ell)(\epsilon[K_{\gamma\beta}]e^{\gamma}_{\mu}e^{\beta}_{\nu}+n_{\mu}n_{\nu}[K]), \label{eq29a} \\
	\mathfrak{R}=\mathfrak{R}^{\pm}-\delta(\ell)2\epsilon[K], \label{eq29b} \\
	T_{\mu\nu}=T_{\mu\nu}^{\pm}+\delta(\ell)S_{\mu\nu}, \label{eq29c}
\end{eqnarray} 
where, $e^{a}_{b}$ is the projection operator from 4-dimensional space-time to 3-dimensional space-time and the normalisation condition reads $n_{a}n^{a}=\epsilon$. Moreover, $\epsilon=\pm1$ for time-like and space-like normalisation. Projection of eq.~\eqref{eq3} in 3-dimensional space-time leads to 
\begin{equation}
	{S}^{\gamma}_{\beta}=-\frac{\epsilon}{8\pi} \Big[{K}^{\gamma}_{\beta}+(2k\eta-1){\delta}^{\gamma}_{\beta}{K}^{\kappa}_{\kappa}\Big], \label{eq29d}
\end{equation} 
For gravastar $\epsilon=1$ and in the context of Rastall theory of gravity, the Lanczos equation is modified as: 
\begin{equation}
	{S}^{\gamma}_{\beta}=-\frac{1}{8\pi} \Big[{K}^{\gamma}_{\beta}+(2k\eta-1){\delta}^{\gamma}_{\beta}{K}^{\kappa}_{\kappa}\Big], \label{eq29e}
\end{equation} 
where, $\eta=0$ retains the general form of the Lanczos equations in Einstein gravity and ${K}_{\gamma\beta}={K}^{+}_{\gamma\beta}-{K}^{-}_{\gamma\beta}$ ,where the $(+)$ and $(-)$ signs indicate the value of $K_{\gamma\beta}$ at the exterior and interior interfaces, respectively. The form of second fundamental ${K}_{\gamma\beta}^{\pm}$ is given below:
\begin{equation}
	{K}_{\gamma\beta}^{\pm}=-{\eta}_{\tau}^{\pm} \Big(\frac{\partial^2 {X_{\tau}}}{\partial{\zeta^{\gamma}} \partial{\zeta^{\beta}}}+{\Gamma}_{\alpha\beta}^{\tau} \frac{\partial{X_{\alpha}}}{\partial {\zeta^{\gamma}}} \frac{\partial{X_{\beta}}}{\partial{\zeta^{\beta}}}\Big). \label{eq30}
\end{equation}
The double sided normal on the surface is defined as, 
\begin{equation}
	\eta_{\tau}^{\pm}={\pm} \Big|{g}^{\alpha\beta} \frac{\partial{f}}{\partial{x^{\alpha}}} \frac{\partial{f}}{\partial{x^{\beta}}}\Big|^{-\frac{1}{2}}\frac{\partial{f(r)}}{\partial{x^{\tau}}}, \label{eq31}
\end{equation}
with $\eta^{\tau}\eta_{\tau}=1$. Following the Lanczos equation \cite{Lanczos}-\cite{Musgrave}, the surface stress-energy tensor at the boundary of the interface is defined as, ${S}_{\gamma\beta}=diag(-\Sigma, \mathfrak{P}, \mathfrak{P}, \mathfrak{P} )$, where $\Sigma$ and $\mathfrak{P}$ are the surface energy density and the surface pressure respectively and are expressed as:
\begin{eqnarray}
	\Sigma=-\frac{1}{4\pi R}{\Big(\sqrt{f(r)}\Big)}^{+}_{-}~, \label{eq32} \\
	\mathfrak{P}=-\frac{\Sigma}{2}-\frac{(2k\eta-1)}{16\pi} \Big(\frac{f'(r)}{\sqrt{f(r)}}\Big)^{+}_{-}, \label{eq33}
\end{eqnarray}
where, $f(r)^{+}_{-}$ represents the $g_{rr}$ components of the exterior and interior regions respectively. Using eqs.~\eqref{eq18} and \eqref{eq28}, eqs.~(\ref{eq32}) and \eqref{eq33} reduce to, 
\begin{equation}
	\Sigma=\frac{1}{4\pi R}\Bigg[\sqrt{1+\frac{c_{1}R^2}{3}+\frac{kq^2}{8\pi R^2}}-\sqrt{1-\frac{2M}{R}+\frac{q^2}{R^2}}\Bigg] , \label{eq34} \\
\end{equation}
\begin{equation}
	\mathfrak{P}=\frac{1}{8\pi R}\Bigg[\frac{1-\frac{M}{R}}{\sqrt{1-\frac{2M}{R}+\frac{q^2}{R^2}}}-\frac{1+\frac{2}{3}c_{1}R^2}{\sqrt{1+\frac{1}{3}c_{1}R^2-\frac{kq^2}{8\pi R^2}}}\Bigg]-\frac{2k\eta}{16\pi R}\Bigg[(\frac{\frac{2M}{R}-\frac{2q^2}{R^2}}{\sqrt{1-\frac{2M}{R}+\frac{q^2}{R^2}}}-\frac{\frac{kq^2}{4\pi R^2}+\frac{2c_{1}R^2}{3}}{\sqrt{1+\frac{1}{3}c_{1}R^2-\frac{kq^2}{8\pi R^2}}})\Bigg].\label{eq35}
\end{equation}
The additional term on the right hand side of eq.~\eqref{eq35} arises due to the modification of Lanczos equation in Rastall theory of gravity. It should be noted that for $\eta=0$, we recover the generalised form of surface pressure in the framework of Einstein gravity. The mass of the gravastar shell is obtained as: 
\begin{equation}
	M_{shell}=4\pi R^2\Sigma=R\Bigg(\sqrt{1+\frac{c_{1}R^2}{3}+\frac{kq^2}{8\pi R^2}}-\sqrt{1-\frac{2M}{R}+\frac{q^2}{R^2}}\Bigg). \label{eq36}
\end{equation} 
The total mass of the gravastar in terms of the mass of the shell can be written as:
\begin{equation}
	M=\frac{4M_{shell}\sqrt{6\pi(24\pi R^2+8c_{1}\pi R^4-3kq^2)}-8c_{1}\pi R^4+24\pi q^2+3kq^2-24\pi M_{shell}^2}{48\pi R}. \label{eq37}
\end{equation}
\section{Boundary Condition}\label{sec5} Analysis of the physical behaviour of this model of charged gravastar depends on the constants that arise through mathematical formalism. In this section, we analyse the boundary conditions by matching the interior region and the thin shell at the interface $r=r_{1}$ and similarly matching the thin shell and the exterior region at the boundary $r=r_{2}$. By matching the solutions, we obtain the numerical values of the necessary constants, namely, $c_{1},c_{3},c_{4}$ and $c_{5}$. Here, we make use of  eq.~\eqref{eq34} and the concept that at the boundary $r=r_{2}$, the surface energy density vanishes \cite{Bcp}. \\
(i) At $r=r_{1}$ we match the interior and thin shell solutions: 
\begin{eqnarray}
	1+\frac{1}{3}c_{1}r_{1}^2-\frac{k}{8\pi}\frac{q^2}{r_{1}^2}=\frac{(4\zeta-1)^2}{(6\zeta-1)(2\zeta-1)}\frac{q^2}{r_{1}^2}+2\frac{(4\zeta-1)}{2\zeta-1}\ln(r_{1})+c_{4}. \label{eq38}\\ \nonumber\\
	c_{3}\Big(1+\frac{1}{3}c_{1}r_{1}^2-\frac{k}{8\pi}\frac{q^2}{r_{1}^2}\Big)=e^{2(\frac{(6-16\zeta)\ln(r_{1})+(2\zeta-1)\ln[q^2(1-4\zeta)+r_{1}^2(6\zeta-1)]}{(8\zeta-2)}+c_{5})}. \label{eq39}
\end{eqnarray}
(ii) At $r=r_{2}$, we match the thin shell and exterior solutions:
\begin{eqnarray}
	\frac{(4\zeta-1)^2}{(6\zeta-1)(2\zeta-1)}\frac{q^2}{r_{2}^2}+2\frac{(4\zeta-1)}{2\zeta-1}\ln(r_{2})+c_{4}=\Big(1-\frac{2M}{r_{2}}+\frac{q^2}{r_{2}^2}\Big). \label{eq40} \\ \nonumber\\
	e^{2(\frac{(6-16\zeta)\ln(r_{2})+(2\zeta-1)\ln[q^2(1-4\zeta)+r_{2}^2(6\zeta-1)]}{(8\zeta-2)}+c_{5})}=\Big(1-\frac{2M}{r_{2}}+\frac{q^2}{r_{2}^2}\Big). \label{eq41}
\end{eqnarray}
We consider the interior radius $r_{1}=10$~Km \cite{Bcp} and accounting for the infinitesimally small thickness of the shell, we take the range of outer radius as $r_{2}=10.001-10.009$~Km. Within the characteristic limit of Rastall theory, we consider the value of Rastall parameter to be $\zeta=0.23$. The prediction \cite{Bcp} that the surface energy density vanishes at the outer radius of the gravastar leads to the numerical evaluation of the necessary constants. Here, we have taken the mass of the gravastar $M=2.75~M_{\odot}$ and the notion of $\frac{2M}{r}<\frac{8}{9}$ is suitably obeyed. It is also noted that this arbitrary combination of radii and mass provide a unique solution as the amount of charge varies. We tabulate the numerical values of all the constants in tab~\ref{tab1}.
\begin{table}[htbp]
	\centering
	\begin{tabular}{|c|c|c|c|}
		\hline
		q & $c_{1}$ & $c_{4}$ & $c_{5}$ \\
		\hline
		0 & -0.0243 & -0.4925 & 20.275 \\
		\hline
		2 & -0.0233 & -0.451 & 20.345 \\
		\hline 
		4 & -0.0205 & -0.3275 & 20.476 \\
		\hline 
		6 & -0.0158 & -0.122 & 20.567 \\
		\hline
	\end{tabular}
	\caption{Numerical evaluation of necessary constants with the variation of charge q. \label{tab1}}
\end{table}
\section{Important features of gravastar in this model} \label{sec6} The finite slice of thickness of the thin shell, the extremely high density $(\gtrsim10^{15}g/cm^{3})$ in this region and through the use of table~\ref{tab1}, we have analysed the physical attributes of a gravastar, {\it viz}, proper length, energy, entropy and the EoS parameter of the shell. 
\subsection{Proper Length of the shell} The thickness of the thin shell is infinitesimally small $(\epsilon=r_{2}-r_{1}<<1)$ \cite{Mazur}-\cite{Mazur2}, and it is the partition between the interior and the exterior region. Using eq.~\eqref{eq25}, the expression of the proper length is obtained as:
\begin{equation}
\ell=\int_{r_{1}}^{r_{2}} e^{\lambda}~dr=\int_{r_{1}}^{r_{2}}\frac{dr}{\sqrt{\frac{(4\zeta-1)^2}{(6\zeta-1)(2\zeta-1)}\frac{q^2}{r^2}+2\frac{(4\zeta-1)}{(2\zeta-1)}\ln(r)+c_{4}}}. \label{eq42}
\end{equation}  
Using eq.~\eqref{eq42} and table~\ref{tab1}, we graphically represent the proper length of the shell with shell thickness $(\epsilon)$. 
\begin{figure}[htbp]
	\centering
	\includegraphics[width=.6\textwidth]{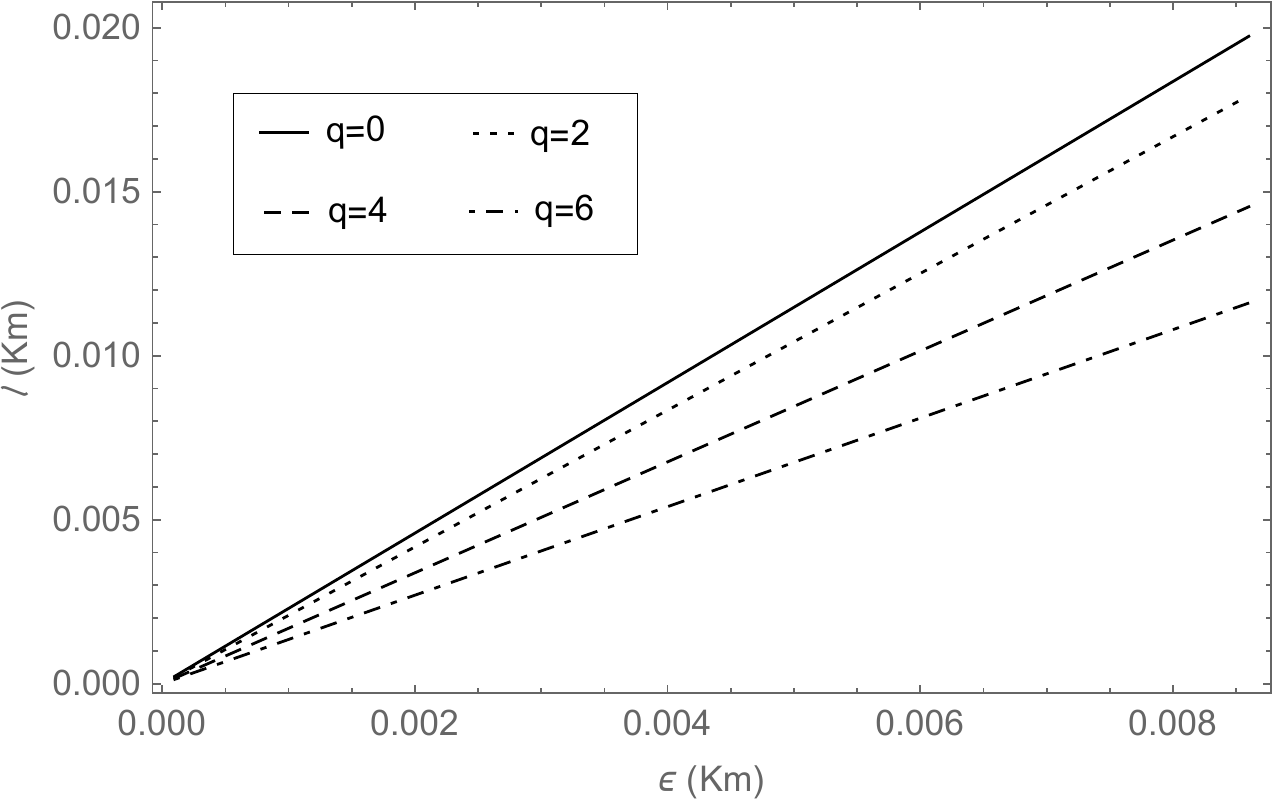}
	\caption{Variation of proper length of the shell $\ell$~(Km) against shell thickness $\epsilon$~(Km).\label{fig2}}
\end{figure}
It is noted here that as the thickness of the shell increases, the proper length also increases. On the other hand, an increase in charge results in a decrease in the proper length. Similar results are also obtained in Refs. \cite{Yousaf,Bhar1}.
\subsection{Energy of the shell} The shell region is characterised by the EoS $p=\rho$, which is the special class of barotropic EoS $p=\omega\rho$, with $\omega=1$. Maur-Mottola in their model hypothesised that the thin shell contains ultra-relativistic stiff fluid which satisfies Zel'dovich's criteria \cite{Zeldovich,Zeldovich1}. The energy contained in the case of a charged gravastar is expressed as:
\begin{equation}
	\mathcal{E}=\int_{r_{1}}^{r_{2}} 4\pi r^2\rho~dr. \label{eq43}
\end{equation}  
Using eq.~\eqref{eq26a} in eq.~\eqref{eq43} we obtain the mathematical form of the energy contained in the shell. Due to the complexity of the equation, we have opted for the graphical representation of the energy of a charged gravastar. 
\begin{figure}[htbp]
	\centering
	\includegraphics[width=.6\textwidth]{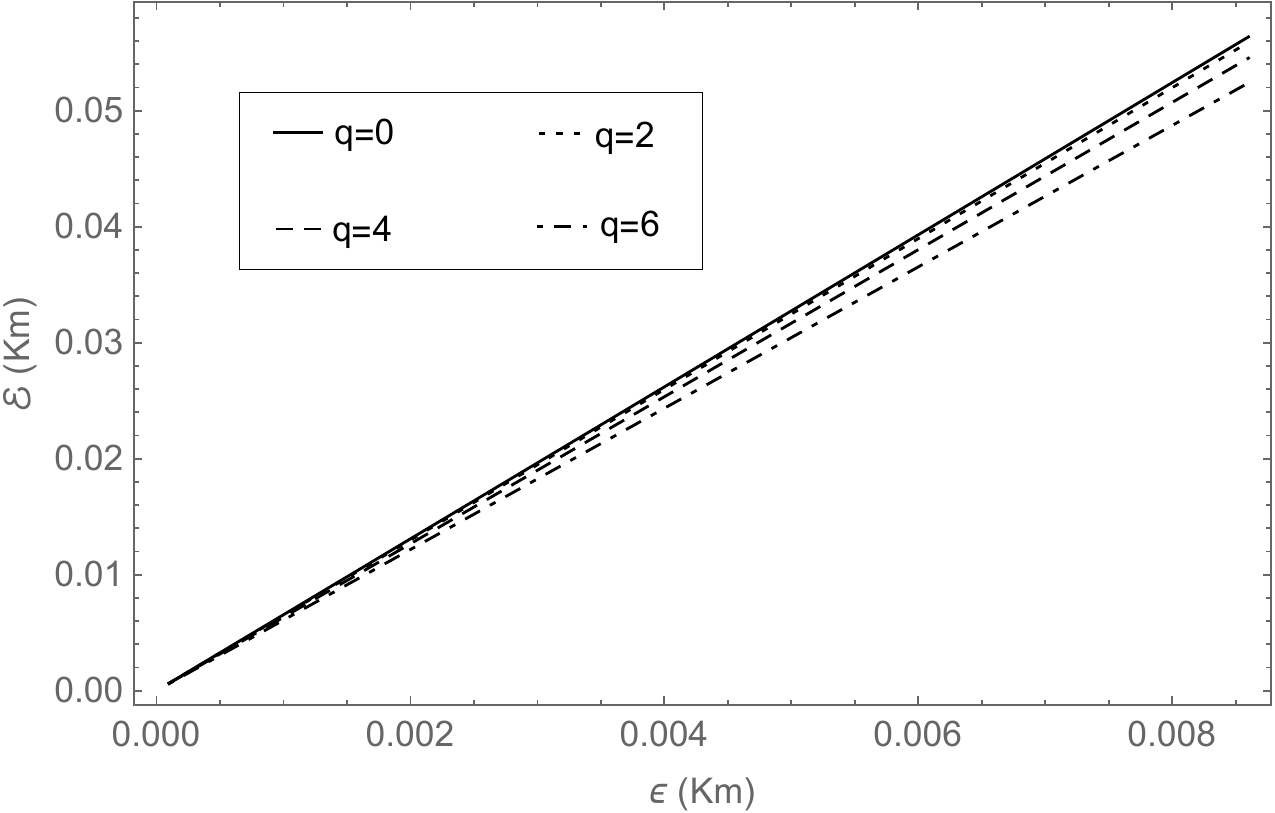}
	\caption{Variation of energy of the shell $\mathcal{E}$~(Km) against shell thickness $\epsilon$~(Km).\label{fig3}}
\end{figure}
From figure~\ref{fig3}, it is evident that more charged fluid is found toward the outer edge of the shell. Moreover, an increase in charge results in a decrease in energy. Similar result has been obtained in the study of Bhar and Rej \cite{Bhar1}. 
\subsection{Entropy of the shell} 
\begin{figure}[htbp]
	\centering
	\includegraphics[width=.6\textwidth]{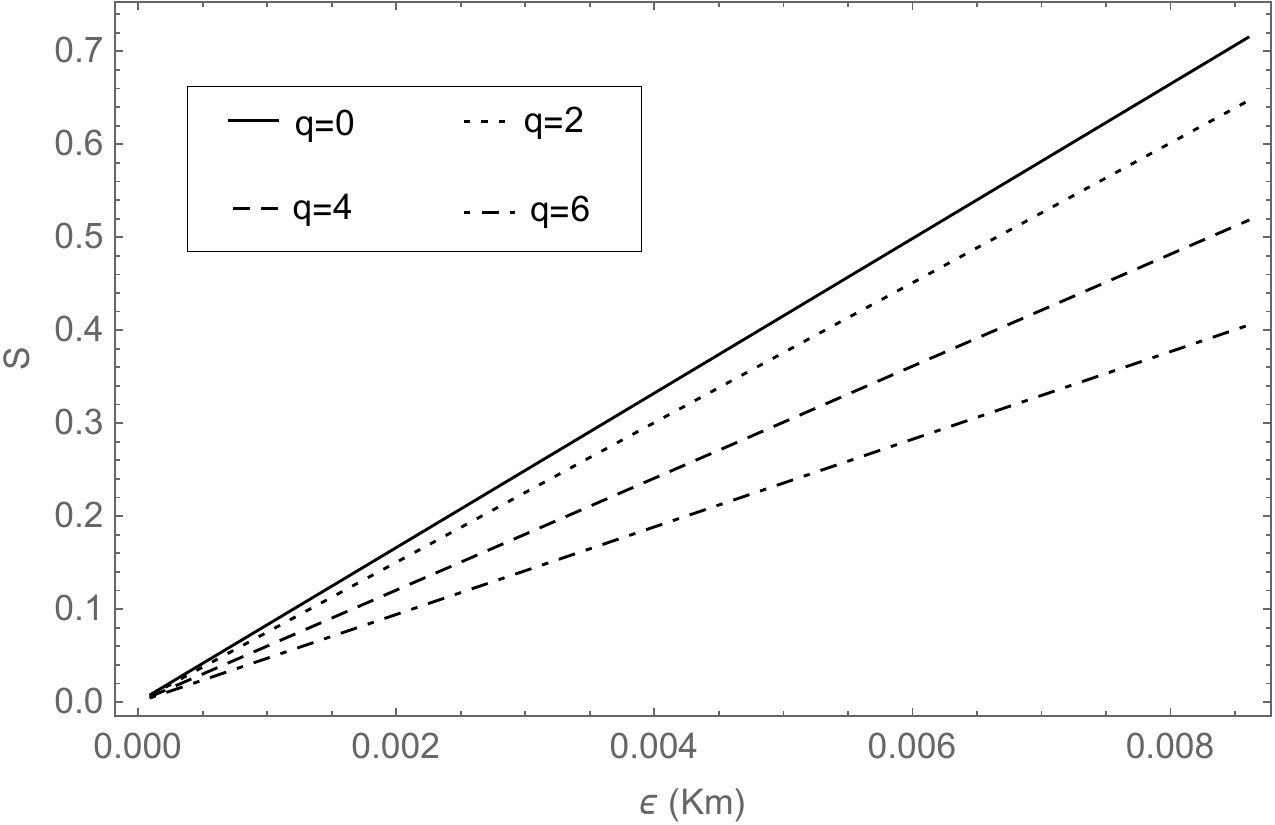}
	\caption{Variation of entropy of the shell $S$ against shell thickness $\epsilon$~(Km).\label{fig4}}
\end{figure}

Entropy is a measure of disorderness in mechanical systems. According to the Mazur-Mottola model \cite{Mazur}-\cite{Mazur2}, the entropy of the vacuum interior is zero, and the thin shell is solely responsible for the entropy considerations. It is evaluated on the basis of the entropy function of the form:
\begin{equation}
	S=4\pi\int_{r_{1}}^{r_{2}} \mathfrak{s}(r)r^2e^{\alpha} dr. \label{eq44}
\end{equation}
Here $\mathfrak{s}(r)=\frac{\alpha^2k_{B}^2T(r)}{4\pi\hbar^2}=\frac{\alpha k_B}{\hbar}\sqrt{\frac{p}{2\pi}}$ is the entropy density corresponding to a local specific temperature $T(r)$. $\alpha$ is a dimensionless constant and we consider $\alpha=1$ without any loss of generality, $k_{B}$ is the Boltzmann constant and $\hbar=\frac{h}{2\pi}$ is the Planck constant. Using Eq.~(\ref{eq26a}) in eq.~\eqref{eq44} we obtain the total entropy in the Planckian units $(\hbar=k_{B}=1)$ as: 
\begin{equation}
	S=2\sqrt{2\pi}\int_{r_{1}}^{r_{2}}\frac{r^{2}\sqrt{\rho_{0}e^{\frac{2(8\zeta-3)\ln(r)+(1-2\zeta)\ln{[q^2(1-4\zeta)+r^2(6\zeta-1)]}}{(2\zeta-1)}}}}{\sqrt{\frac{(4\zeta-1)^2}{(6\zeta-1)(2\zeta-1)}\frac{q^2}{r^2}+2\frac{(4\zeta-1)}{(2\zeta-1)}\ln(r)+c_{4}}}~dr. \label{eq45}
\end{equation}
The variation of entropy of the thin shell with the variation of charge and shell thickness is shown in figure~\ref{fig4}.
A linear relationship between the entropy and shell thickness is obtained, and it is also noted that as the charge increases, the entropy decreases. Usmani et al. \cite{Usmani} showed that the entropy of a charged gravastar depends on the thickness of the shell, which is also found to be true in our model. On the other hand, a similar result was also obtained by Yusaf et al. \cite{Yousaf}. 
\subsection{EoS parameter} The EoS parameter at the hypersurface $r=R$ is expressed as:
\begin{equation}
	\mathcal{W}=\frac{\mathfrak{P}}{\Sigma}. \label{eq46}
\end{equation}
Substituting the values of $\Sigma$ and $\mathfrak{P}$ from eqs.~\eqref{eq34} and \eqref{eq35} in eq.~\eqref{eq46}, we obtain:
\begin{equation}
	\mathcal{W}(R)=\frac{\Bigg[\frac{1-\frac{M}{R}}{\sqrt{1-\frac{2M}{R}+\frac{q^2}{R^2}}}-\frac{1+\frac{2}{3}c_{1}R^2}{\sqrt{1+\frac{1}{3}c_{1}R^2-\frac{kq^2}{8\pi R^2}}}\Bigg]}{2\Bigg[\sqrt{1+\frac{c_{1}R^2}{3}+\frac{kq^2}{8\pi R^2}}-\sqrt{1-\frac{2M}{R}+\frac{q^2}{R^2}}\Bigg]}-\frac{k\eta}{2}\frac{\Bigg[\frac{\frac{2M}{R}-\frac{2q^2}{R^2}}{\sqrt{1-\frac{2M}{R}+\frac{q^2}{R^2}}}-\frac{\frac{kq^2}{4\pi R^2}+\frac{2c_{1}R^2}{3}}{\sqrt{1+\frac{1}{3}c_{1}R^2-\frac{kq^2}{8\pi R^2}}}\Bigg]}{\Bigg[\sqrt{1+\frac{c_{1}R^2}{3}+\frac{kq^2}{8\pi R^2}}-\sqrt{1-\frac{2M}{R}+\frac{q^2}{R^2}}\Bigg]}. \label{eq47}
\end{equation}
To obtain a real solution of $\mathcal{W}(R)$, we impose the restriction $\frac{2M}{R}-\frac{q^{2}}{R^2}<1$ or $2M<(R+\frac{q^{2}}{R})$ which is in essence satisfied by eq.~\eqref{eq28}. The sign of $\mathcal{W}(R)$ depends on the sign of numerator or denominator. Interestingly, for large values of $R$, $\mathcal{W}(R)\rightarrow-1+\zeta$, where $\zeta=k\eta$ is the Rastall parameter and for $\zeta=0$, the EoS steps into the dark energy EoS with $\mathcal{W}(R)\rightarrow-1$. On the other hand, for small values of $R$, $\mathcal{W}(R)\rightarrow -\zeta$. Consequently, for $\zeta=0$,  $\mathcal{W}(R)\rightarrow 0$ which defines a dust shell.

\section{Stability Analysis} \label{sec7} This section is devoted to the stability analysis of the present theoretical model. The two relevant methods employed to study the stability of gravastar in this model are i) Surface Redshift and ii) Entropy maximisation. 
\subsection{Surface Redshift} 
\begin{figure}[htbp]
	\centering
	\includegraphics[width=.6\textwidth]{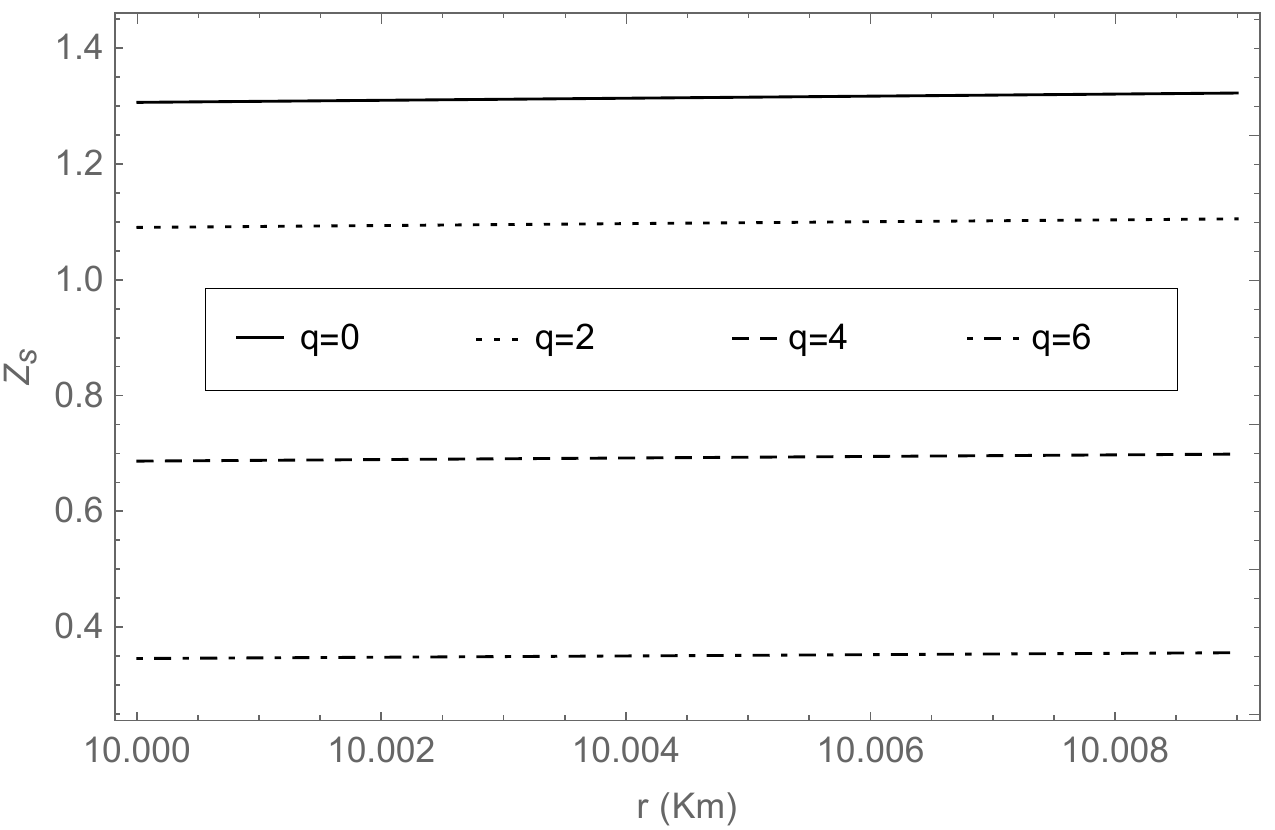}
	\caption{Radial variation of surface redshift $(Z_{s})$.\label{fig5}}
\end{figure}

Investigation of gravitational surface redshift can be considered a key ingredient in analysing the stability of a gravastar. In fact, this study can be useful for obtaining information regarding the detection of gravastars. The surface redshift is expressed as $Z_{s}=\frac{\lambda_{e}-\lambda_{o}}{\lambda_{e}}$, where $\lambda_{o}$ and $\lambda_{e}$ are the observed and emitted signals, respectively. Buchdahl asserted that for a static, isotropic perfect fluid distribution, the value of the surface redshift should not exceed $2$, i.e., $Z_{s}<2$ \cite{Buchdahl}-\cite{Bohmer}. Barraco and Hamity \cite{Barraco} further stated that $Z_{s}<2$ is also a valid condition in the absence of cosmological constant. But B$\ddot{o}$hmer and Harko \cite{Bohmer} showed that the presence of anisotropy in the stellar configuration extends the surface redshift to $Z_{s}\leq5$. Ivanov in his study of anisotropic stars \cite{Ivanov} further extended the limit of surface redshift from $Z_{s}<3.84$ to $Z_{s}\leq5.211$. DeBenedictis et al. \cite{Debenedictis} examined the stability of gravastars through axial perturbations and studied the surface redshift parameter in gravastars. Their study revealed that the surface redshift was consistent with the above discussion. \\
The surface redshift is expressed as:
\begin{equation}
	Z_{s}=-1+\frac{1}{\sqrt{g_{tt}}}. \label{eq48}
\end{equation}  
From eq.~\eqref{eq26} and $g_{tt}=e^{2\nu}$ we obtain:
\begin{equation}
	Z_{s}=-1+\frac{1}{\sqrt{e^{2\Big(\frac{(6-16\zeta)\ln(r)+(2\zeta-1)\ln[q^2(1-4\zeta)+r^2(6\zeta-1)]}{(8\zeta-2)}+c_{5}\Big)}}}. \label{eq49}
\end{equation}  
We have represented the radial variation of surface redshift in figure~\ref{fig5}. We have found that the $Z_{s}$ increases with the shell thickness and decreases with the increment of charge. 
In light of the above discussion, we may claim that our present model of charged gravastar is stable and physically viable.  
\subsection{Entropy maximisation} In this section, we have used the same method as employed by Mazur and Mottola in their original work \cite{Mazur}-\cite{Mazur2} to check the stability criteria of gravastar model. To ensure a stable model, we have applied the procedure of maximisation of entropy functional with respect to the mass function $M(r)$ and checked the sign of the second variation of the entropy functional. The first variation must vanish at the thin shell boundaries at $r=r_{1}$ and $r=r_{2}$, i.e., $\delta S=0$. The entropy functional is written as:
\begin{equation}
	S=\frac{\sqrt{2}\alpha k_{B}}{2\hbar}\int_{r_{1}}^{r_{2}}rdr\Big(\frac{dM}{dr}\Big)^{\frac{1}{2}}\frac{1}{\sqrt{g_{rr}}}, \label{eq50}
\end{equation} 
where, $\alpha$ is a dimensionless constant, $k_{B}$ is Boltzmann constant, $\hbar=\frac{h}{2\pi} $ is the Planck constant and $g_{rr}=1-\frac{2M}{r}+\frac{q^{2}}{r^{2}}$. Using eqs.~\eqref{eq25}, \eqref{eq28} and \eqref{eq38}, we can obtain the mass $M(r)$ in Rastall gravity in the form:
\begin{equation}
	M=-r\Bigg[\frac{(4\zeta-1)^{2}}{2(2\zeta-1)(6\zeta-1)}q^{2}(\frac{1}{r^{2}}-\frac{1}{r_{1}^{2}})+\frac{(4\zeta-1)}{(2\zeta-1)}\ln(\frac{r}{r_{1}})-\frac{q^{2}}{r^{2}}-\frac{kq^{2}}{16\pi r_{1}^{2}}+\frac{c_{1}r^{2}}{6}\Bigg].\label{eq51}
\end{equation}  
Therefore, following Refs.\cite{Mazur}-\cite{Mazur2} we can obtain the second variation of entropy functional in the form:
\begin{equation}
	\delta^{2}S=\frac{\sqrt{2}\alpha k_{B}}{2\hbar} \int_{r_{1}}^{r_{2}}rdr\Big(\frac{dM}{dr}\Big)^{-\frac{3}{2}}\frac{1}{\sqrt{g_{rr}}}\Bigg[ -\Bigg(\frac{d(\delta M)}{dr}\Bigg)^{2}+\frac{(\delta M)^{2}}{g^{2}r^{2}}\Bigg(\frac{dM}{dr}\Bigg)\Bigg(1+\frac{dM}{dr}\Bigg)\Bigg]. \label{eq52}
\end{equation}
Here we have conveniently considered a linear combination of $M(r)$ as $\delta M=\chi\eta$, where $\eta$ vanishes at the boundaries. Plugging this consideration in eq.~\eqref{eq52} and integrating by parts keeping $\delta M=0$ at the apex of the shell we obtain:
\begin{equation}
	\delta^{2}S=-\frac{\sqrt{2}\alpha k_{B}}{2\hbar}\int_{r_{1}}^{r_{2}}rdr\Big(\frac{dM}{dr}\Big)^{-\frac{3}{2}}\frac{1}{\sqrt{g_{rr}}}\chi^{2}\Big[\frac{d\psi}{dr}\Big]^{2}<0. \label{eq53}
\end{equation}
It is evident that the entropy function in Rastall gravity attains a maximal value for all the radial variations which vanish at the extreme points of the thin shell of the charged gravastar. Therefore, we can conclusively say that the perturbations in the ultra-relativistic fluid in the region of thin shell further decrease the entropy of the region, which in turn proves that our model of charged gravastar is stable against small perturbations having fixed end points. Fundamentally, the employment of modified gravity theory does not affect the stability of a charged gravastar.    
\section{Conclusion}\label{sec8} In this paper, we have investigated the gravastar formalism in the framework of Rastall theory of gravity in spherically symmetric space-time and studied the effect of charge in the formulation of gravastar. We have obtained a new class of exact, singularity free and physically acceptable solutions for three layers of an isotropic, charged gravastar, and we have noted the interesting features that appear in the configuration of gravastar due to the incorporation of Rastall theory. The following prominent gravastar attributes are listed below:
\begin{itemize}
	\item {\bf Interior solution:} Using the EoS $p=-\rho$ and the conservation of the energy-momentum tensor equation eq.~\eqref{eq16}, it is found that the matter density and pressure are constant in the interior region. In this region, we have obtained singularity free solutions as described in eqs.~\eqref{eq18} and \eqref{eq19}. Additionally, we have determined the analytical expression of the active gravitational mass as given in eq.~\eqref{eq20} contained in the interior region of a charged gravastar. 
	\item {\bf EoS of the thin shell:} Eq.~\eqref{eq26a} shows the behaviour of matter density as well as the pressure of the thin shell region, and it is illustrated graphically in figure~\ref{fig1}. The figure depicts the presence of denser fluid toward the outer edge of the shell.  
	\item {\bf Junction Condition:} We have modified the junction condition in the framework of Rastall gravity and found that for null Rastall parameter, we retain the junction conditions of Einstein gravity. The interior and exterior solutions are matched at the smooth hypersurface $r=R$ following Israel condition \cite{Israel,Israel1}. Through the matching condition, we have obtained the intrinsic surface energy tensor given in eq.~\eqref{eq29}. The formulation of Lanczos equation \cite{Lanczos}-\cite{Musgrave} yields the surface energy density and surface pressure, which are used to express the thin shell mass in terms of the total mass of the gravastar as given in eq.\eqref{eq36}. 
	\item {\bf Proper length of the thin shell:} Figure~\ref{fig2} shows the radial variation of proper length $(\ell)$ of the thin shell of a charged gravastar. It is evident that the proper length increases linearly with shell thickness and the increase in charge results in a decrease of length.  
	\item{\bf Energy of the shell:} Figure~\ref{fig3} shows the radial variation of the energy of the charged fluid contained within the shell. With increasing shell thickness, more charged fluid is found, and when the charge increases the energy decreases which in turn illustrates the non-repulsive nature of the energy in the shell. 
	\item{\bf Entropy:} In figure~\ref{fig4} we find the increasing nature of the entropy of a charged gravastar. It is also noted that fluid contains less entropy in presence of electric charge than uncharged fluid. 
	\item{\bf EoS parameter:} The formulation of the EoS parameter of the shell region shows that for large values of radius $(R)$, the EoS indicates the dark energy regime $(\mathcal{W}(R)\rightarrow-1)$ when Rastall parameter $\zeta\rightarrow0$. On the other hand, for small values of $R$, $\mathcal{W}(R)\rightarrow -\zeta$. Consequently, for $\zeta\rightarrow0$, $\mathcal{W}(R)\rightarrow 0$ which defines a dust shell configuration.
	\item {\bf Surface Redshift:} The stability of the model is ensured through the consideration of gravitational surface redshift which is found to obey the Buchdahl limit \cite{Buchdahl} for isotropic stellar configurations. 
	\item {\bf Entropy Maximisation:} The fulfillment of the entropy maximisation condition shows that our model of gravastar is stable under small radial perturbations in presence of charge also. The entropy function attains a maximum value as evident from eq.~\eqref{eq53} when radial perturbations vanish at the end point of the shell. 
\end{itemize}   
From the above discussion, we may conclude that charge has some significant effect on the formation and basic properties of gravastar and charged gravastar models may exist in Rastall theory of gravity. One of the most noteworthy features of this model is that for $q=0$, we recover the results which are similar in form to those obtained by Ghosh et al. \cite{Bcp}. We have found that the use of Rastall theory modifies the form of the basic properties of a charged gravastar from those obtained via Einstein gravity. Interestingly, equating Rastall parameter to zero $(\zeta=0)$ in this model retains Einstein's general relativity formalism. Finally, our model represents a stable and viable stellar configuration admitting all the characteristic features of a stable charged gravastar model within the framework of Rastall theory. 
\acknowledgments
DB is thankful to DST for providing the fellowship vide no: DST/INSPIRE/03/2022/000001. We are also thankful to Jo\~ao Luis Rosa, Assistant Professor, University of Gdansk, Poland for the fruitful discussions on our manuscript.   

%\paragraph{Note added.} This is also a good position for notes added
%after the paper has been written.

% The bibliography will probably be heavily edited during typesetting.
% We'll parse it and, using the arxiv number or the journal data, will
% query inspire, trying to verify the data (this will probalby spot
% eventual typos) and retrive the document DOI and eventual errata.
% We however suggest to always provide author, title and journal data:
% in short all the informations that clearly identify a document.

\end{document}